\documentclass{aa}

\usepackage{graphicx}
\usepackage{txfonts}

\usepackage{booktabs}
\usepackage{natbib}
\usepackage{mathtools}
\usepackage{ amssymb }
\usepackage[usenames,dvipsnames]{color} 
\usepackage{xspace}
\usepackage[flushleft]{threeparttable}
\usepackage[normalem]{ulem}

\usepackage[colorlinks = true,
            linkcolor = blue,
            urlcolor  = blue,
            citecolor = blue,
            anchorcolor = blue]{hyperref}

\def\aa{$\alpha\alpha$\xspace}
\def\ga{$\gamma\alpha$\xspace}

\begin{document}

   \title{Populations of double white dwarfs in Milky Way satellites and their detectability with LISA}

   \author{V. Korol \inst{1} 
          \and S. Toonen \inst{1}        
          \and A. Klein \inst{1}
          \and V. Belokurov \inst{2}
          \and F. Vincenzo \inst{1,3}
          \and R. Buscicchio \inst{1}
          \and D. Gerosa \inst{1}
          \and C.~J. Moore \inst{1}
          \and \\E. Roebber \inst{1}
          \and E.~M. Rossi \inst{4}
          \and A. Vecchio \inst{1}
          }

   \institute{Institute for Gravitational Wave Astronomy, School of Physics and Astronomy, University of Birmingham, Birmingham, B15 2TT, UK  
    \email{korol@star.sr.bham.ac.uk, toonen@star.sr.bham.ac.uk}
    \and Institute of Astronomy, Madingley Rd, Cambridge CB3 0HA, United Kingdom
    \and Center for Cosmology and AstroParticle Physics, The Ohio State University, 191 West Woodruff Avenue, Columbus, OH 43210, USA
    \and Leiden Observatory, Leiden University, PO Box 9513, 2300 RA, Leiden, the Netherlands\\
}

   \date{Received February 18, 2020; accepted  May 7, 2020}

  \abstract
   {Milky Way dwarf satellites are unique objects that encode the early structure formation and therefore represent a window into the high redshift Universe. So far, their study has been conducted using electromagnetic waves only. The future Laser Interferometer Space Antenna (LISA) has the potential to reveal Milky Way satellites  through gravitational waves emitted by double white dwarf (DWD) binaries.}
   {We investigate gravitational wave (GW) signals that will be detectable by LISA as a possible tool for the identification and characterisation of the Milky Way satellites.}
   {We used the binary population synthesis technique to model the population of DWDs in dwarf satellites and we assessed the impact on the number of LISA detections when making changes to the total stellar mass, distance, star formation history, and metallicity of satellites. We calibrated predictions for the known Milky Way satellites on their observed properties.}
   {We find that DWDs emitting at frequencies $\gtrsim 3\,$mHz can be detected in Milky Way satellites at large galactocentric distances. The number of these high frequency DWDs per satellite primarily depends on its mass, distance, age, and star formation history, and only mildly depends on the other assumptions regarding their evolution such as metallicity. We find that dwarf galaxies with $M_\star>10^6\,$M$_{\odot}$ can host detectable LISA sources; the number of detections scales linearly with the satellite's mass. We forecast that out of the known satellites, Sagittarius, Fornax, Sculptor,  and  the  Magellanic Clouds can be detected with LISA.}
   {As an all-sky survey that does not suffer from contamination and dust extinction, LISA will provide observations of the Milky Way and dwarf satellites galaxies, which will be valuable for Galactic archaeology and near-field cosmology.} 


   \maketitle

\section{Introduction}

Dwarf galaxies are the first baryonic systems to appear in the Universe.
The $\Lambda\,$Cold Dark Matter ($\Lambda$CDM) cosmological model predicts that dwarf galaxies develop from small fast-growing lumps of dark matter  that are able to accrete and cool down enough gas to form stars. 
State-of-the-art $\Lambda$CDM cosmological dark matter-only simulations predict the existence of a large number of small dark matter halos, which are large enough to host dwarf galaxies, within the Milky Way-like halos \citep[e.g.][]{spr08,kuh09,sta09,gar14}. 
Over time, a fraction of these dwarf galaxies are destroyed by the gravitational pull of the Milky Way and now form the diffuse halo of our Galaxy \citep[e.g.][]{bul05}. Those that survived are known as Milky Way satellites.
Both satellites and stars of the diffuse halo bear the Galactic archaeological record and are,  therefore, precious tools to reconstruct the Milky Way formation history.

About 60 satellite galaxies orbiting within the virial radius of the Milky Way are known to date, with stellar masses reaching down to $\sim300\,M_\odot$ \citep[e.g. ][]{sim19}.
This census is known to be highly incomplete \citep[e.g.][]{Koposov2009,new18}.
Modern surveys such as the Sloan Digital Sky Survey \citep[SDSS, see e.g.][]{Willman2005,Belokurov2007,Torrealba2016} and the Dark Energy Survey \citep[DES, e.g.][]{Koposov2015,bec15,drl15} do not cover the entirety of the sky and are subject to detectability limits that depend on both the surface brightness of and the distance to the satellites \citep[see][]{Koposov2008,Tollerud2008}.
Even the upcoming Rubin Observatory Legacy Survey of Space and Time \citep[][]{ive08} would only be able to find about half of the expected 
dwarfs galaxies \citep{Jethwa2018,new18}.

In a companion paper by \citet{roe20}, we show that the Laser Interferometer Space Antenna (LISA) can detect ultra-short period ($\lesssim 10$\,min) double white dwarfs (DWDs) that are hosted in Milky Way satellites, and that using the LISA data we can associate them to the host. Therefore, LISA will provide a complementary survey to study the satellites. In this paper we investigate the properties of DWD populations in satellites and how they are affected by, for example, star formation history (SFH) and metallicity ($Z$) of the satellites. Among the variety of stellar-mass binaries that will be observable by LISA, we focus on DWDs because they are expected to be the most numerous GW sources for a given galaxy mass \citep[e.g.][]{Nel01d,bre19}.

The intrinsic faintness of white dwarf stars in the optical band limits their detection to distances of only a few kiloparsecs even for state-of-the-art facilities; the farthest known DWD is at $\sim$2.5\,kpc \citep{bur19}. Thus, no confirmed detections outside the Milky Way are known to date. The only exception is the X-ray source, RX J0439.8-6809, that has been tentatively identified as an accreting white dwarf in a compact binary system with a Helium white dwarf donor in the Large Magellanic Cloud \citep{1994A&A...281L..61G,1997A&A...323L..41V}. However, later spectral modelling suggests that this binary is also consistent with an unusually hot white dwarf in the Milky Way halo \citep{2015A&A...584A..19W}. 

A number of theoretical studies predict that tens of thousands of Milky Way DWDs should be detectable by LISA \citep{Nel01,Ruit10,Kor17,bre19,Lamberts19}; outside the Galaxy, DWDs can be detected out to the edge of the Local Group, and specifically up to a few tens of sources should be observed in the Andromeda galaxy \citep{kor18}. Although black hole and neutron star binaries are stronger GW emitters in the LISA band compared to DWDs, their rates are expected to be at least three orders of magnitude lower in the Milky Way and only a few sources are predicted to be detectable in nearby massive satellites \citep[e.g. ][]{and19,lau19,sesa19,set19}.

Leveraging large cosmological simulations, \cite{Lamberts19} have shown that DWDs can indeed be detected by LISA in both satellites and tidal stellar streams.
Crucially, the overall number of detections depends on the detailed properties of the considered halo and its accretion history.
Cosmological simulations consider global solutions but the specific distribution of observed Milky Way satellites is not reproduced yet.
In this paper we take a pragmatic approach to estimate the DWD population in Galactic satellites. We simulate individual DWD populations tuned on the properties of different satellites and investigate the expected number of LISA detections as a function of total stellar mass, distance, SFH, and metallicity of the satellites. By directly calibrating predictions on the observed properties of the known Milky Way satellites, our approach allows us to draw solid forecasts on the expected number of detections for the LISA mission. More importantly, we provide estimates for the range of expected detections and demonstrate how they depend on the main astrophysical processes at work. This allow us to make conclusions on additional and/or complementary information that LISA observations could offer for studying Milky Way satellites.

The outline of this manuscript is as follows. In Section \ref{sec:method} we describe our DWD population synthesis procedure as well as our dwarf-galaxy model. In Section \ref{sec:results} we report our results for a generic satellite and also present the number of predicted detections in the currently known Milky Way satellite population. In Section \ref{sec:discussion} we compare DWDs against electromagnetic (EM) mass-tracers and discuss how DWD observations with LISA can be used to measure the satellite properties. Finally, we summarise our results in Section \ref{sec:conclusions}.

\section{Method} \label{sec:method}

In this study we perform binary population synthesis to assess the prospects of GW detections in Milky Way dwarf satellites. 
Calculations are performed using the publicly available code \texttt{SeBa} \citep{Por96, Too12} that models the formation of DWDs starting from Zero-age Main-sequence (ZAMS) stars. Synthetic catalogues of DWDs produced with \texttt{SeBa} have been carefully calibrated against state-of-the-art observations of DWDs in terms of both mass ratio distribution \citep{Too12} and number density \citep{Too17}. We construct different models at different metallicity, SFH, and treatment of the unstable mass-transfer phase (the so-called common envelope, CE). Each \texttt{SeBa} model consists of $3\times 10^6$ binaries at birth that roughly correspond to $10^7\,M_\odot$ in total. 

\subsection{Initial binary population}

We initialise the progenitor binary population as follows.
Studies of Galactic globular clusters, the Magellanic Clouds (LMC, SMC), and local dwarf spheroidal galaxies find that the resolved stellar mass function appears to be consistent with that observed in the field populations and young forming clusters of the Milky Way \citep[cf.][]{bas10}.
We adopt the  \cite{Kro93} initial mass function (IMF) for the mass of the primary  star, $m_1$.
In this work, the primary (secondary) star is considered to be the initially most (least) massive star of the pair. 
We simulate primary stars in the mass range $m_1\in [0.95\,M_\odot, 10\,M_\odot]$. To calculate how many DWDs are present in a satellite of a given mass, we consider that stellar masses range from $0.1-100\,M_\odot$.
The mass of the secondary star $m_2$ is drawn uniformly in $[0.08\,M_\odot, m_1]$ \citep{Rag10, Duc13}.

Semi-major axes $a$ are drawn from a distribution that is uniform in $\log(a)$ \citep{Abt83}. We consider only detached binaries on the ZAMS with orbital separations up to $10^6\,R_\odot$.
Eccentricities $e$ are initialised according to a thermal distribution in $[0,1]$ \citep{Heg75}. 
Local dwarf galaxies show a diversity in metallicities ranging from roughly 0.0001 to Solar metallicity \citep[see e.g.][and references therein]{mcc12}. We adopt three different values: $Z=0.0001,0.001$, and $0.02$, where the default is $Z=0.001$.
We set a constant binary fraction of $50\%$. This is appropriate for typical white dwarf progenitor (A- to F-type stars) at Solar metallicity, but it likely underestimates the multiplicity of early B-type stars \citep[][]{derosa2014, Duc13, Moe17}.

\subsection{Binary evolution}

The progenitor of a DWD with orbital period $P<$ a few hours typically undergoes  two phases of mass transfer  taking place when each of the stars evolves off the main sequence. Typically, at least one of these mass transfer phases is unstable and leads to a CE surrounding the binary \citep{Paczynski1976,Webbink1984}. 
The CE evolution takes place on the timescale of thousand years, during which one of the two stars expands and engulfs the companion causing both objects to orbit inside a single, shared envelope \citep{Ivanova2013}.
The companion star spirals inwards through the envelope, losing energy and angular momentum due to the dynamical friction. The temperature of the envelope consequently increases.
This phase continues until the envelope is ejected from the system leaving behind the core of the expanded star and its companion  in a tighter orbit. 

We adopt two different treatments for the CE phase: the $\alpha$-formalism based on the energy conservation and the $\gamma$-formalism based on the balance of angular momentum \citep[for a review see][]{Ivanova2013}.
More specifically, following \citet{Too12} we make use of two evolutionary models, that we denote \aa  and \ga. 
In model \aa,  the $\alpha$-formalism is used to determine the outcome of every CE. For model \ga the $\gamma$-prescription is applied unless the binary contains a compact object or the CE is triggered by a tidal instability, in which case $\alpha$-prescription is used.
For a typical evolution of a system according to the \ga model, the first CE is typically described by the $\gamma$ formalism, while the second by the $\alpha$ formalism. 
We highlight that the \ga model is specifically calibrated for DWDs trough a reconstruction of the evolutionary paths of individual observed binaries \citep{Nel00,nelemans05,van06}.

Our treatment of CE evolution has an effect on the  DWD  formation rate. In particular, model \ga predicts about twice as many DWDs compared to model \aa \citep{Too17}. When restricting to those with orbital periods accessible to LISA, the two models are more similar. The predicted number of visible LISA sources in the Milky Way varies by $\lesssim 25\%$ \citep[][see also Section~\ref{sec:results} of this paper]{Kor17}.  

\begin{figure}
	\centering
	\includegraphics[width=\columnwidth]{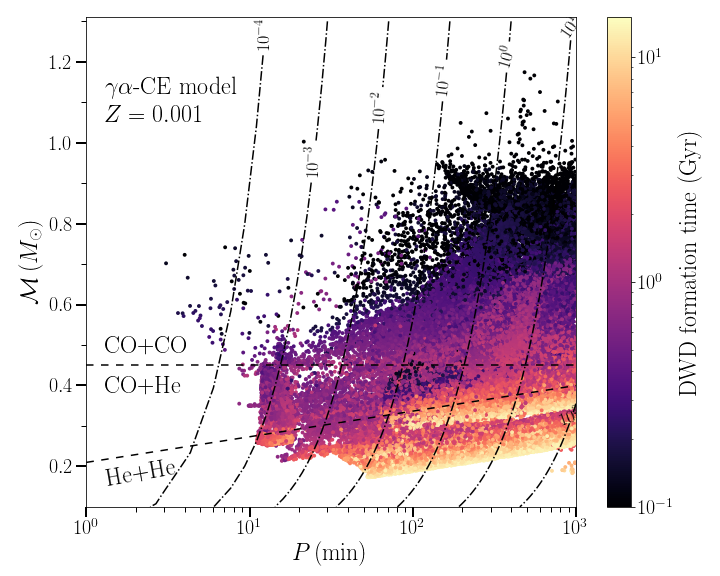}
	\caption{Orbital periods $P$ and chirp mass ${\cal M}$ of DWDs at birth. The DWD formation time is represented by the colour scale, while their merger time by  GW emission is represented by dashed-dotted contours (both timescales are expressed in Gyr). The figure shows only a fraction of the population with orbital periods accessible to LISA. Dashed lines represent approximate boundaries between DWDs of different types: CO+CO, CO+He and He+He; ONe white dwarfs represent a negligible fraction of the population and occupy the same part of the parameter space as CO+He ones. We also note that DWDs in this figure do not encode any information about the type of the host galaxy.} 
	\label{fig:pop0}
\end{figure}

Figure~\ref{fig:pop0} illustrates the synthetic population obtained by running \texttt{SeBa} with our fiducial assumptions 
and the CE $\gamma \alpha$-model. All binaries are initialised at the same time and evolved until both stars turn into white dwarfs.
The $x$ and $y$ axes show orbital period $P$ and chirp mass ${\cal M}= (m_1 m_2)^{3/5}/(m_1+m_2)^{1/5}$ of DWDs at birth, meaning that they do not encode any information about the type of the host galaxy. We describe how we transform these properties to account for the age and the SFH of the satellite galaxy in Section~\ref{sec:sfh}. The colour scale indicates their formation time measured from ZAMS. Dashed-dotted contours represent the binary merger time assuming that it is driven by GW radiation only \citep[e.g. ][]{Maggiore}.
Dashed lines  delineate approximate boundaries between different core compositions of white dwarf components in our data (Carbon-oxygen, CO; or Helium, He). 
Figure~\ref{fig:pop0} includes a number of features related to the choices made in binary population synthesis (for example a cutoff for DWDs with ${\cal M} \lesssim 0.4\,M_\odot$ and $P \lesssim10$\,min). These choices and their effects are fully detailed in \citet{Too12}.

DWDs tend to occupy the lower-right part of the parameter space and accumulate at long orbital periods and chirp masses of $\sim 0.5\,M_\odot$.
The formation time increases with decreasing chirp mass:
in particular, at most a few Gyr are necessary to form more massive CO+CO and CO+He DWDs, while the formation of He+He systems takes several Gyr. 
The sum of the formation and merger times roughly indicates the lifespan of the binary. For instance, DWDs with formations times $\lesssim 1$~Gyr (darker in Fig.~\ref{fig:pop0}) and merger timescales $\lesssim 10^{-3}\,$Gyr (top-left in Fig.~\ref{fig:pop0}) are short-lived binaries and would generally inhabit star-forming environments. 
On the contrary, yellow circles in the bottom-right region of Fig.~\ref{fig:pop0} require a longer time to form and merge, and  would typically be present in old ($\gtrsim 10$~Gyr) stellar populations. The age and SFH of the satellite play a crucial role in determining the properties of the resulting DWDs.

\subsection{Synthetic satellites}

Using the terminology of \citet{bul17}, systems with    stellar mass $M_\star<10^9\,M_\odot$ are referred to as `dwarf' galaxies.
Dwarfs are further subdivided into `bright' dwarfs  ($10^7M_\odot  < M_\star < 10^9 M_\odot$), `classical' dwarfs with ($10^5M_\odot < M_\star < 10^7 M_\odot$), and `ultra-faint' dwarfs ($10^2 M_\odot < M_\star < 10^5 M_\odot$). 
More specifically, we model satellites with masses $M_\star\in[10^3 M_\odot-10^{10}M_\odot]$,  covering from ultra-faint dwarfs to Large Magellanic Cloud-analogues.

\subsubsection{Star formation histories} \label{sec:sfh}

The availability of detailed colour-magnitude diagrams for an increasing number of Local Group galaxies revealed that these systems have diverse SFHs, ranging from dwarfs dominated by old stars ($\gtrsim 12$\,Gyr ago) to nearly constantly star forming environments \citep{tol09,bro14,wei14,wei19}.
It is generally found that the SFH depends on both the satellite's mass and  morphological type (spheroidal, elliptical, irregular, transitional). In particular, ultra-faint dwarf galaxies form 80\% of their stellar mass by redshift $z\sim2$, while bright dwarfs produce only 30\% of their stars by the same time \citep{wei14}.
This trend becomes more complicated if one considers the dwarf's environment. Ultra-faint galaxies are generally found within the virial radius of the Milky Way, and thus experienced processes like tidal interactions and ram-pressure stripping that are known to quench star formation \citep[e.g. ][]{wet15,fil18}. In contrast, bright dwarfs are typically located outside the sphere of gravitational influence of the Milky Way, and are thus less likely to be influenced by the environment.
Similar trends have been also observed in numerical simulations, such as APOSTLE \citep{saw16} and Auriga \citep{gra16}: galaxies with $10^5 < M_\star/ M_\odot < 10^6$ tend to have declining SFHs, massive dwarfs $10^7 < M_\star/M_{\odot} < 10^9$ show an increasing star formation peaking at recent times, and the intermediate cases are found to form stars at a roughly constant rate \citep{dig19}.

To cover such a variety of cases we adopt three simple SFH models:  single burst, constantly star-forming, and exponentially declining (also known as $\tau$-model, \citealt{bru83}). We adopt the exponentially declining model as our fiducial SFH model.
For simplicity we assume that all satellites have the same age of 13.5\,Gyr.

\begin{figure}
	\centering
	\includegraphics[width=1\columnwidth]{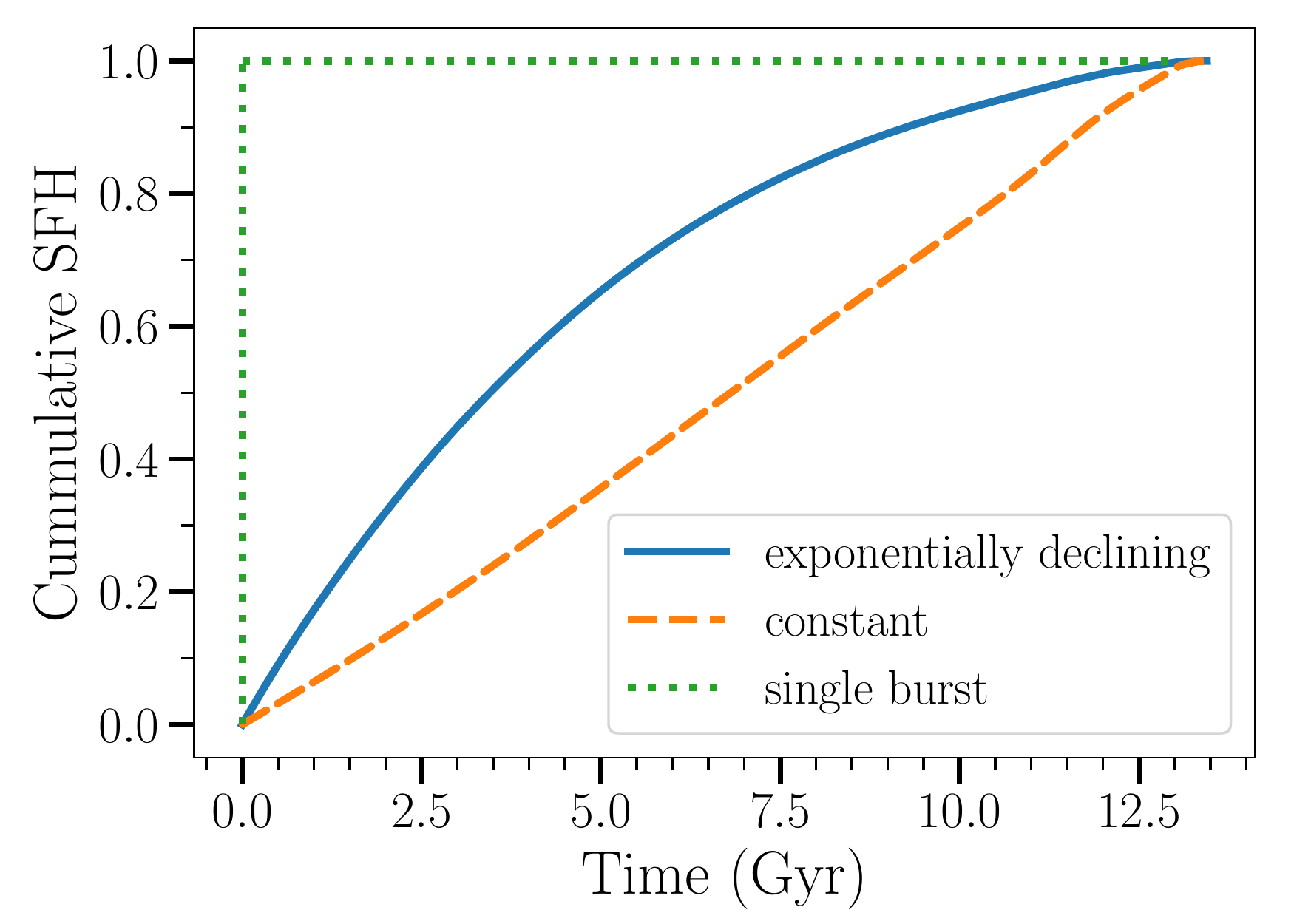}
	\caption{Adopted SFHs: exponentially declining (blue solid), constant (orange dashed) and single burst  13.5\,Gyr ago (green dotted).} 
	\label{fig:SFHs}
\end{figure}

These different SFHs are implemented in our study as follows.
For the single burst scenario, we model the DWD orbital evolution from their formation until 13.5\,Gyr in which their orbits shrink through GW radiation reaction. We discard binaries if their formation times greater than 13.5\,Gyr, they have begun mass transfer (i.e. when one of the white dwarfs fills its Roche lobe) or they have already merged within this time. 
These manipulations affect mostly short-lived systems and deplete the central part of Fig.~\ref{fig:pop0}. The constant SFH is produced by injecting DWD binaries into a satellite (accounting for  the delay time between the parent binary and DWD formation, cf. Fig.~\ref{fig:pop0}) at the rate of $1\,M_\odot$\,yr$^{-1}$ and subsequently evolving their orbits until present time.
Over 13.5\,Gyr, this model produces a galaxy of $13.5 \times 10^{10}\,M_\odot$.
The exponential SFH is constructed in a similar way, but the DWD injection rate decreases according to an exponential with characteristic timescale $\tau = 5\,$Gyr \citep{wei14}. The result is then normalised to a total mass of $10^{10}\,M_\odot$. 
We note that more complex star formation scenarios can be constructed by combining the three basic ones described in this Section.
In Figure~\ref{fig:SFHs} we plot the cumulative distribution of the number of DWDs in the satellite galaxy for different star formation histories as a function of time. 

We evaluate the number of DWDs in a satellite galaxy of mass $M_\star$ by linearly re-scaling the simulation outputs
\begin{equation}
    N_{\rm DWD} = \frac{M_\star}{M_{\rm \texttt{SeBa}}} N_{\rm DWD,\texttt{SeBa}},
\end{equation}
where $N_{\rm DWD,\texttt{SeBa}}$ is the number of DWDs in the synthetic population and where  $M_{ \rm \texttt{SeBa}}$ is the total simulated population mass. 

\subsection{LISA detectability} \label{sec:snr}

LISA is an ESA-lead space mission designed to detect GW sources in the mHz frequency band \citep{ama17}.
The diversity and large amount of GW signals simultaneously present in the LISA data stream make the data analysis extremely challenging \citep[e.g. ][]{bab10, lit20}.
For simplicity, in this paper we use analytic prescriptions to assess the detectability of DWDs with LISA, allowing us to quickly process large populations. A companion paper by \citet{roe20} carefully addresses prospects for detecting and characterising DWDs in Milky Way satellites with LISA.

For a typical DWD the timescale on which the orbital period changes due to the GW emission is significantly longer than the LISA mission lifetime, $T$.
Therefore, one can safely approximate its signal as monochromatic with frequency $f=2/P$. 
Averaging over sky location, polarisation, and inclination, one can write down the signal-to-noise ratio as \cite[e.g.][]{Robson_2019}:
\begin{equation} \label{eqn:snr}
    \rho^2 = \frac{24}{25} |{\cal A}|^2 \frac{T}{S_{\rm{n}}(f) R(f)},
\end{equation} 
where ${\cal A}$ is the amplitude of the signal 
\begin{equation}
    {\cal A} = \frac{2(G{\cal M})^{5/3}(\pi f)^{2/3}}{c^4 d},
\end{equation}
$S_{\rm{n}}(f)$ is the power spectral density (PSD) of the detector noise in the low-frequency limit, $G$ and $c$ are respectively the gravitational constant and the speed of light, and $R(f)$ is a transfer function encoding finite-armlength effects at high frequencies, tending to unity as $f \to 0$.  We compute $R(f)$ numerically by averaging the full time-delay interferometry response of the detector to a monochromatic GW at each frequency as outlined in \citet{Larson_2000}. 
Approximate analytical expressions for $R(f)$ are also given in ~\citet{Robson_2019}\footnote{We note that $R(f)$ in \citet{Robson_2019} is the inverse of our definition.} and~\citet{LISAdoc}.
Both agree well with our numerical result at frequencies below 20~mHz.

\begin{figure}
	\centering
	\includegraphics[width=\columnwidth]{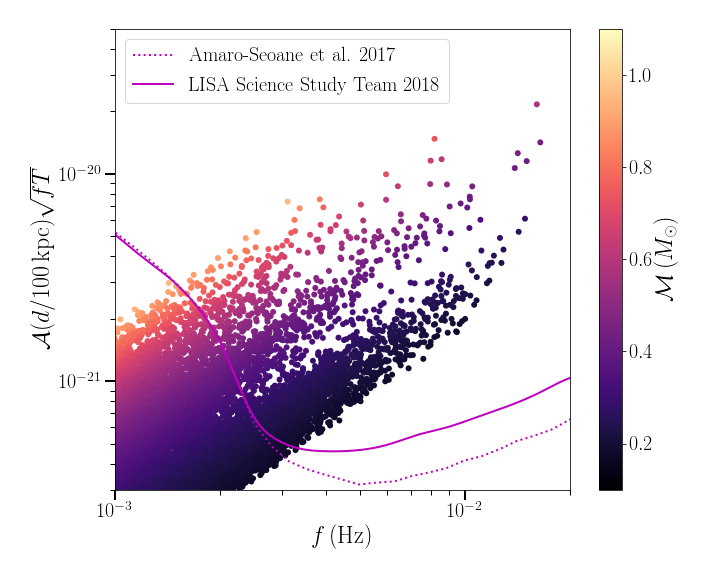}
	\caption{Typical frequencies and characteristic strain for our fiducial population model (coloured circles) placed at $d=100$\,kpc. The colour scale shows the chirp mass distribution.
	Solid and dotted lines indicate the sky-, inclination and polarisation-averaged LISA sensitivity curve of \cite{LISAdoc} and that of  \cite{ama17}, respectively.} 
	\label{fig:amp_def}
\end{figure}

Current LISA specifications \citep{LISAdoc} provide:
\begin{align}
  S_{\rm{n}}(f) &= \frac{1}{L^2} \left[ \frac{4 S_{\rm{acc}}(f)}{(2 \pi f)^4} + S_{\rm{shot}}(f) \right], \\
  S_{\rm{acc}}(f) &= \left( 3 \times 10^{-15} {\rm{~m}}/{\rm{s}}^2 \right)^2 \left[ 1 + \left(\frac{0.4 {\rm{~mHz}}}{f} \right)^2 \right] {\rm{Hz}}^{-1}, \\
  S_{\rm{shot}}(f) &= \left(15 {\rm{~pm}} \right)^2 {\rm{Hz}}^{-1}, \\
  L &= 2.5 {\rm{~Gm}}.
\end{align}
We note that this PSD differs slightly from the one presented in the original LISA mission proposal \citep[][cf. Fig~\ref{fig:amp_def}]{ama17}.
The two PSDs have a different frequency dependence at low frequencies, and are proportional to each other at high frequencies by a factor of 2/3. The updated curve penalises high frequency sources that, as we show later, are accessible at larger distances and therefore are optimal for detecting satellites. Both noise curves account for the confusion foreground noise produced by unresolved Galactic DWDs using the fitting expression of~\cite{Babak:2017tow}.

We adopt the nominal (extended) mission duration time of 4\,yr (10\,yr). That is, we consider a formal duty cycle of 100\% and ignore maintenance operations and data gaps due to, e.g.  laser frequency switches, high-gain antenna re-pointing, orbit corrections, and unplanned events \citep[e.g. ][]{bag19}. 
A more realistic assumption would be to consider a 70-80\% duty cycle as achieved by the LISA Pathfinder mission \citep{LPF2016}, corresponding to $\sim$3\,yr ($\sim$8\,yr) nominal (extended) mission duration. Using Eq.~\eqref{eqn:snr} one can re-scale the signal-to-noise ratio of any individual source (and thus the total number of detections by multiplying the nominal and extended mission results by  $\sqrt{3/4} \simeq 0.87$ and $\sqrt{4/5} \simeq 0.89$, respectively.
A number of studies have assessed the detection threshold for monochromatic sources in the LISA data.  
For example, \cite{CrowderCornish} report a detection threshold of $\rho_{\text{thr}}=5$, while \cite{PhysRevD.81.063008} a threshold of $\rho_{\text{thr}}=5.7$. However, the former study did not include the frequency derivative in the search parameters, and neither one included the second derivative of the frequency or the orbital eccentricity, potentially important parameters to identify mass-transferring or triple systems \citep[e.g. ][]{Nel04,rob18,tam19}. Including those extra parameters would tend to increase the detection threshold. Determining the new threshold would require a study beyond the scope of the present one, we therefore choose a somewhat conservative threshold of $\rho_{\text{thr}}=7$. We verified that increasing the threshold to $\rho_{\text{thr}}=8$ decreases the number of detected binaries by about 20\%.

Figure~\ref{fig:amp_def} shows the sky-, inclination- and polarisation-averaged dimensionless characteristic strain of LISA, $h_{\rm n}= \sqrt{25 f S_{\rm n}(f) R(f) / 24}$ (magenta solid line), and that of our fiducial population $h_{\rm c} = {\cal A}\sqrt{f T}$ at $d=100$\,kpc (coloured circles). 
The fiducial DWD population was obtained starting from binaries represented in Fig.~\ref{fig:pop0} and by computing their properties at the present time assuming that the satellite's age is 13.5\, Gyr and an exponentially declining SFH as described in Section~\ref{sec:sfh}.

In the parameter space represented in Fig.~\ref{fig:amp_def}, the signal-to-noise ratio can be visually estimated as the height above the noise curve \cite[e.g.][]{moo15}. For example, moving same population to $d=1$\,Mpc results into translating all circles down by a factor of 10. Because the distance is fixed, binaries occupy a narrow region in Fig.~\ref{fig:amp_def}, which is set by the minimum  and maximum chirp mass of the population ($0.2\,M_\odot \lesssim {\cal M} \lesssim 1.1\,M_\odot$) as shown in colour. 

\section{Results} \label{sec:results}

\begin{table*}
\centering
\caption{Number of detectable DWDs in a satellite with total stellar mass of $2 \times 10^7\,$M$_\odot$, age of 13.5\,Gyr, and distance of $20$\,kpc for 4-year (10-year) long LISA mission assuming three SFH models: exponentially declining (fiducial), constantly star-forming and single burst. These numbers scales linearly with the satellite's mass. Note also that these results can be simply re-scaled to a different binary fraction by multiplying by 1.3 and 1.6 to get binary fractions of 70\% and 90\% respectively.}
\label{tab:results}
\begin{tabular}{l|c|c|c|ccc}
\hline \hline
CE evolution model & \multicolumn{3}{c|}{$\gamma \alpha$} & \multicolumn{3}{c}{$\alpha \alpha$}                                   \\ \hline
Metallicity  & 0.0001    & 0.001     & 0.002    & \multicolumn{1}{c|}{0.0001} & \multicolumn{1}{c|}{0.001}  & 0.002 \\ \hline 
Exponentially declining SFH & 45  (66) & 37 (58) & 31 (47) & \multicolumn{1}{c|}{41 (56)} & \multicolumn{1}{c|}{35 (48)} & 28 (39) \\Constant SFH    & 53 (98)    & 54 (96)    & 45 (83)    & \multicolumn{1}{c|}{56 (99)} & \multicolumn{1}{c|}{54 (99)} & 46 (84) \\
Single burst SFH & 47 (54)     & 20 (47)     & 26 (52)    & \multicolumn{1}{c|}{27 (34)}  & \multicolumn{1}{c|}{14 (41)}  & 31 (42) \\ \hline 
\end{tabular}
\end{table*}

\begin{figure}
	\centering
    \includegraphics[width=0.78\columnwidth]{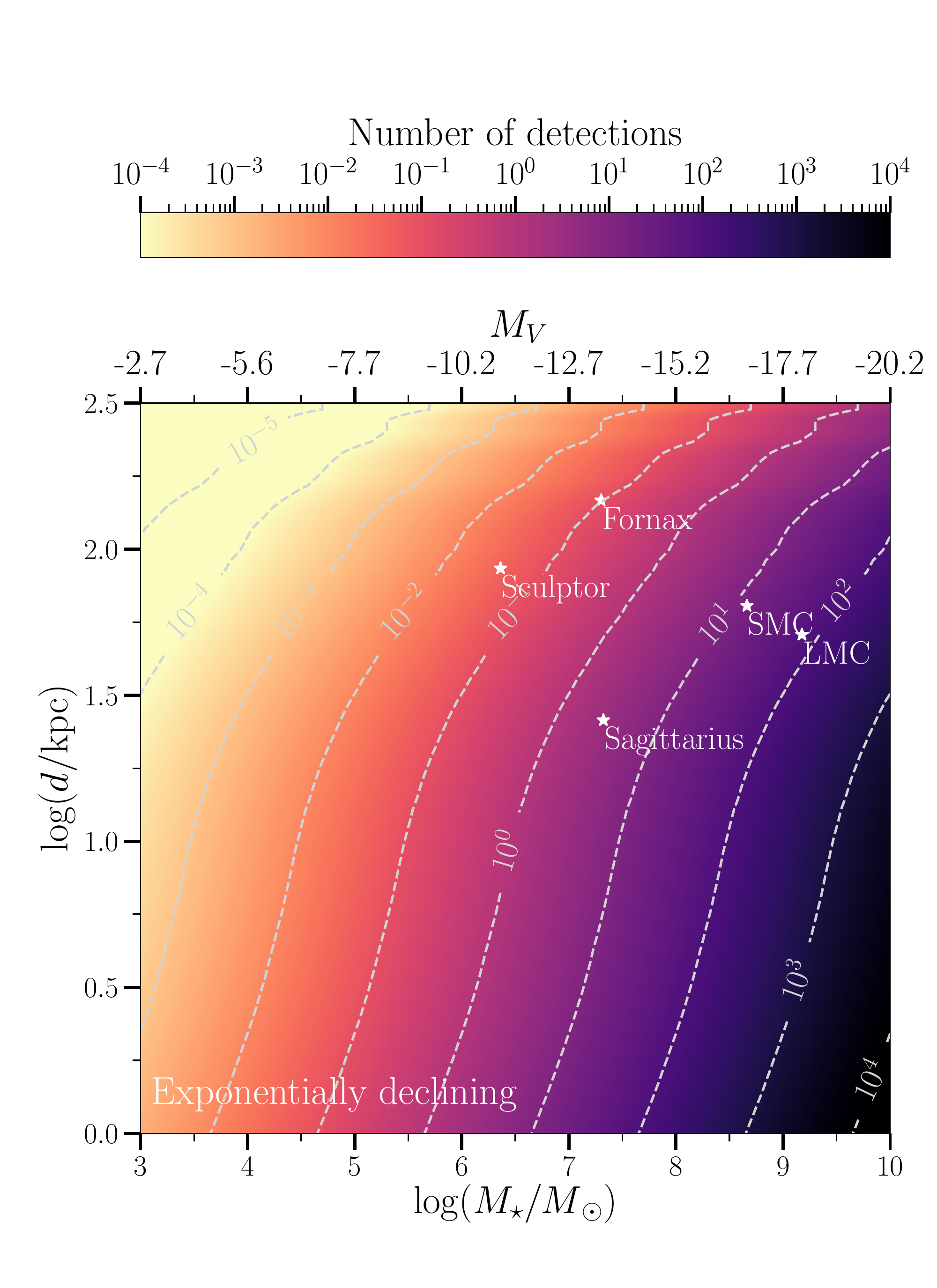}
    \includegraphics[width=0.78\columnwidth]{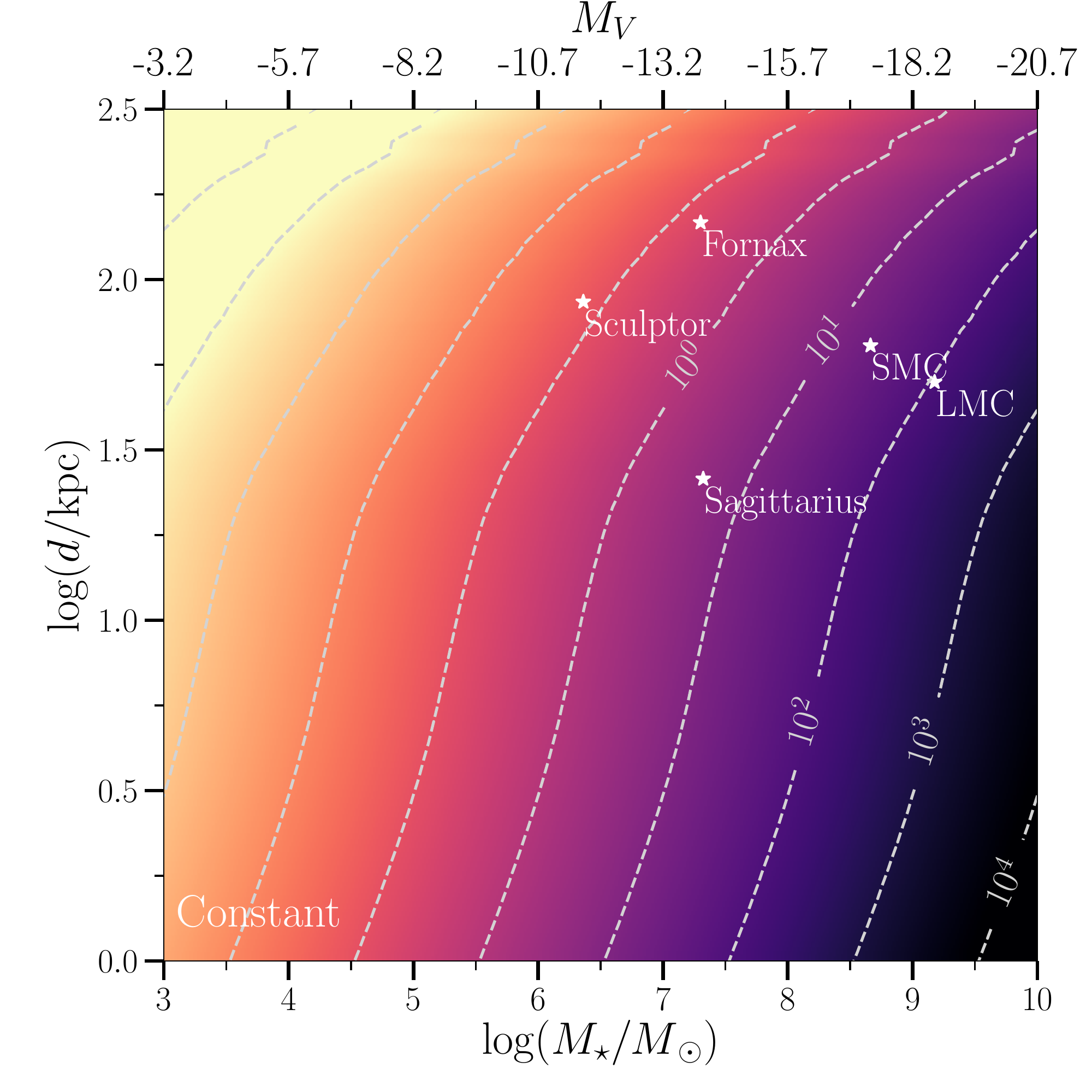}
    \includegraphics[width=0.78\columnwidth]{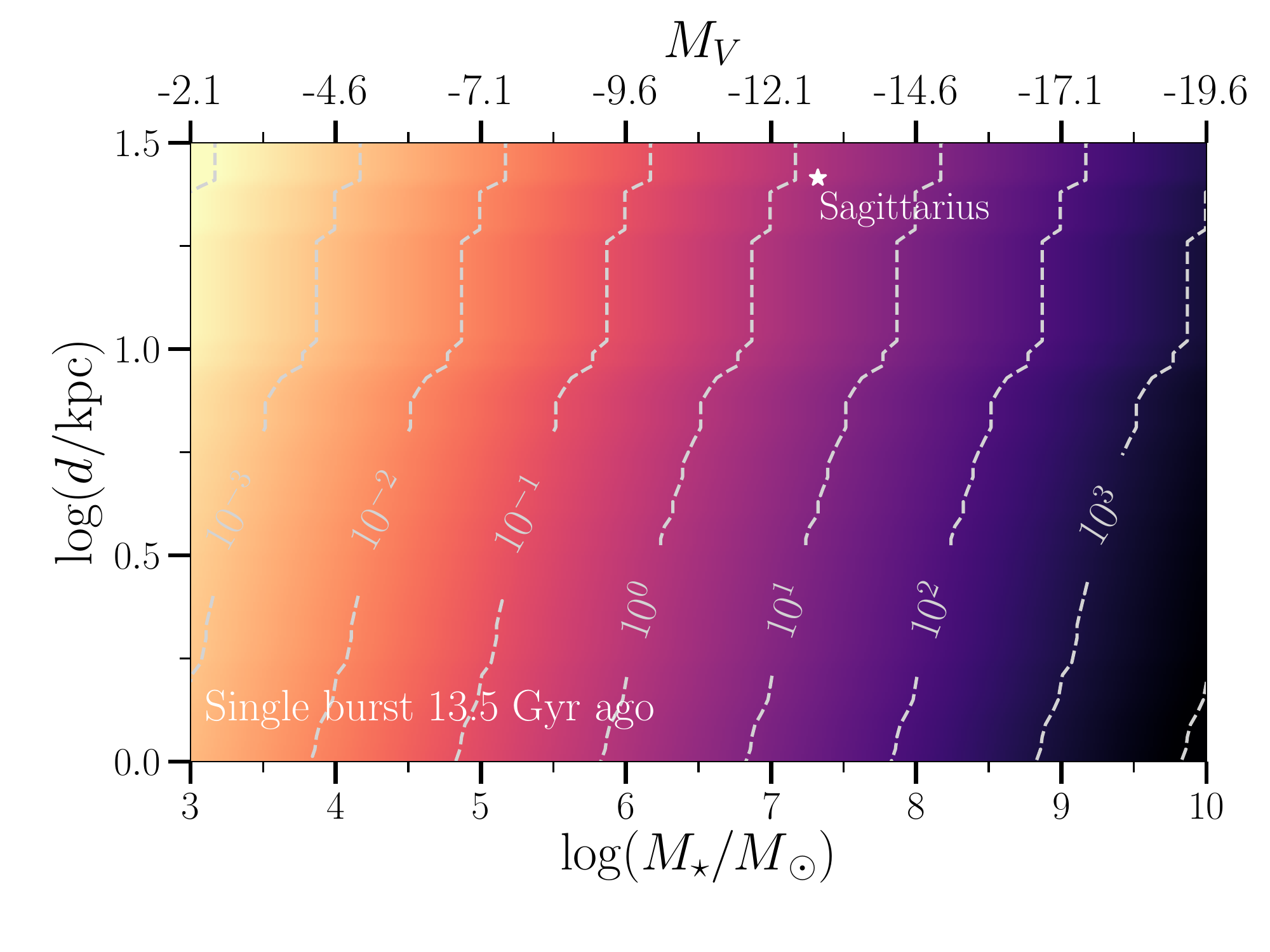}
	\caption{Number of detectable DWDs in the log satellite mass - log distance parameter space for different SFH models:  exponentially decreasing, constantly star-forming at the rate of $1\,M_\odot$\,yr$^{-1}$ and single burst 13.5\,Gyr ago. White stars represent known satellites listed in Tab.~\ref{table:known_satellites}. The top $x$-axes shows the absolute magnitude in the $V-$band. The bottom panel is cut at $\log d \sim$1.6 as there are no more detectable sources at larger distances; this is the case of ultra-faint dwarfs, which have no sources detectable beyond this distance.} 
	\label{fig:detections}
\end{figure}

\subsection{Effect of star formation history}\label{sec:nr}

Here we report the number of DWDs that can be detected by LISA in a given satellite galaxy with a stellar mass $M_\star$ at a distance $d$. Figure~\ref{fig:detections} presents results for our fiducial assumptions ($Z=0.001$, Kroupa IMF, $\gamma \alpha$-CE evolution, and binary fraction of 50\%) and the three different SFH models (exponentially declining, constant, and single burst). For all three SFH models we assume the age of the satellite to be 13.5\, Gyr.

Our exponential SFH model predicts a few million DWDs with orbital frequencies $> 10^{-4}$\,Hz. For a simulated total mass of $10^{10}\,M_\odot$ and assuming  $d=100\,$kpc, we find $\sim$115 ($\sim$294) detectable DWDs in the nominal (extended) LISA mission. 
The number of detections increases linearly with the satellite mass and decreases with its distance.  This is shown in the top panel of Fig.~\ref{fig:detections}. LISA sources are detectable in satellites with $M_\star \gtrsim 10^6\,M_\odot$ \citep[for instance the Sagittarius dwarf spheroidal galaxy,][]{iba94} up to $\sim$30\,kpc, and in Magellanic Cloud-analogues with $M_\star\sim10^9\,M_\odot$ up to $100 - 200\,$kpc (which corresponds to the virial radius of the Milky Way). 
We also find that our model predicts $3.4 \times 10^3$ detections for $M_\star = 2 \times 10^{10}\,M_\odot$ at $d=8.5\,$kpc. This is in agreement with estimates from \citet{kor19} for the MW bulge.  
In order to enable comparison with electromagnetic tracers, Fig.~\ref{fig:detections} shows the absolute $V$-band magnitude of the population, computed using the simple model of \citet{bru03} and the publicly available \texttt{python} package \texttt{smpy}. 

The middle panel of Fig.~\ref{fig:detections} illustrates the number of detections for the constant SFH, keeping all other choices fixed to the fiducial model. In this case, satellites can be detected farther out in the Milky Way halo compared to those in the  exponential SFH model. This is because a constant SFH produces a greater number of DWDs at recent times. We verify that the constant SFH model leads to 7 (51) detections for an Andromeda-like galaxy at the distance of 800\,kpc for the nominal (extended) LISA mission lifetime, in agreement with earlier work of \citet{kor18}.

Ultra-faint dwarfs typically stop forming stars after an initial burst \citep[e.g. ][]{wei14,sim19}.
This scenario is represented in the bottom panel of 
Fig.~\ref{fig:detections}. 
It is evident that ultra-faint dwarfs with $M_{\star}\lesssim10^5 M_\odot$ will be invisible to LISA.
These satellites do not contain DWDs emitting at high frequencies, because they have already long since merged (cf. Fig.~\ref{fig:pop0} for the relevant timescales). 
Only DWDs with frequencies $f>3$\,mHz can be detected and localised at distances $\sim100\,$kpc as a consequence of the fact 
that (i) LISA is maximally sensitive around $3-5\,$mHz (see Fig.~\ref{fig:amp_def}) and (ii) these frequencies are not affected by the confusion foreground \citep[e.g. ][]{lit19}.

Our results illustrate that the total stellar mass sets the fuel supply to generate DWDs, while the SFH together with the age of the satellite determine how many DWDs emit in the LISA frequency band at the present time. In particular, for a fixed satellite mass, age and distance the constant SFH produces on average twice as many detections compared to the exponential model, while the single burst produces only about half. Consequently, all three panels of Fig.~\ref{fig:detections} appear relatively similar when using a  logarithmic scale. 
The crucial difference is given by the number of DWDs with
$f>3$\,mHz hosted by a satellite at the present time.
High frequency DWDs are  more abundant in young and/or star-forming satellites. 
This is because the birthrate of DWDs peaks at early times ($\sim1$\,Gyr), and compact systems merge on shorter timescales (cf. Fig.~\ref{fig:pop0}). This is analogous to the case of massive DWD mergers studies in the context of type Ia-supernovae  \citep[e.g. ][]{Ruit07, Too12, Cla14}. 

We summarise our results for the range of model variations in Table~\ref{tab:results} using the example of a satellite with a total stellar mass of $2 \times 10^7\,$M$_\odot$, an age of 13.5\,Gyr, and a distance of $20$\,kpc.

\subsection{Effect of metallicity}

\begin{figure}
	\centering
	\includegraphics[width=\columnwidth]{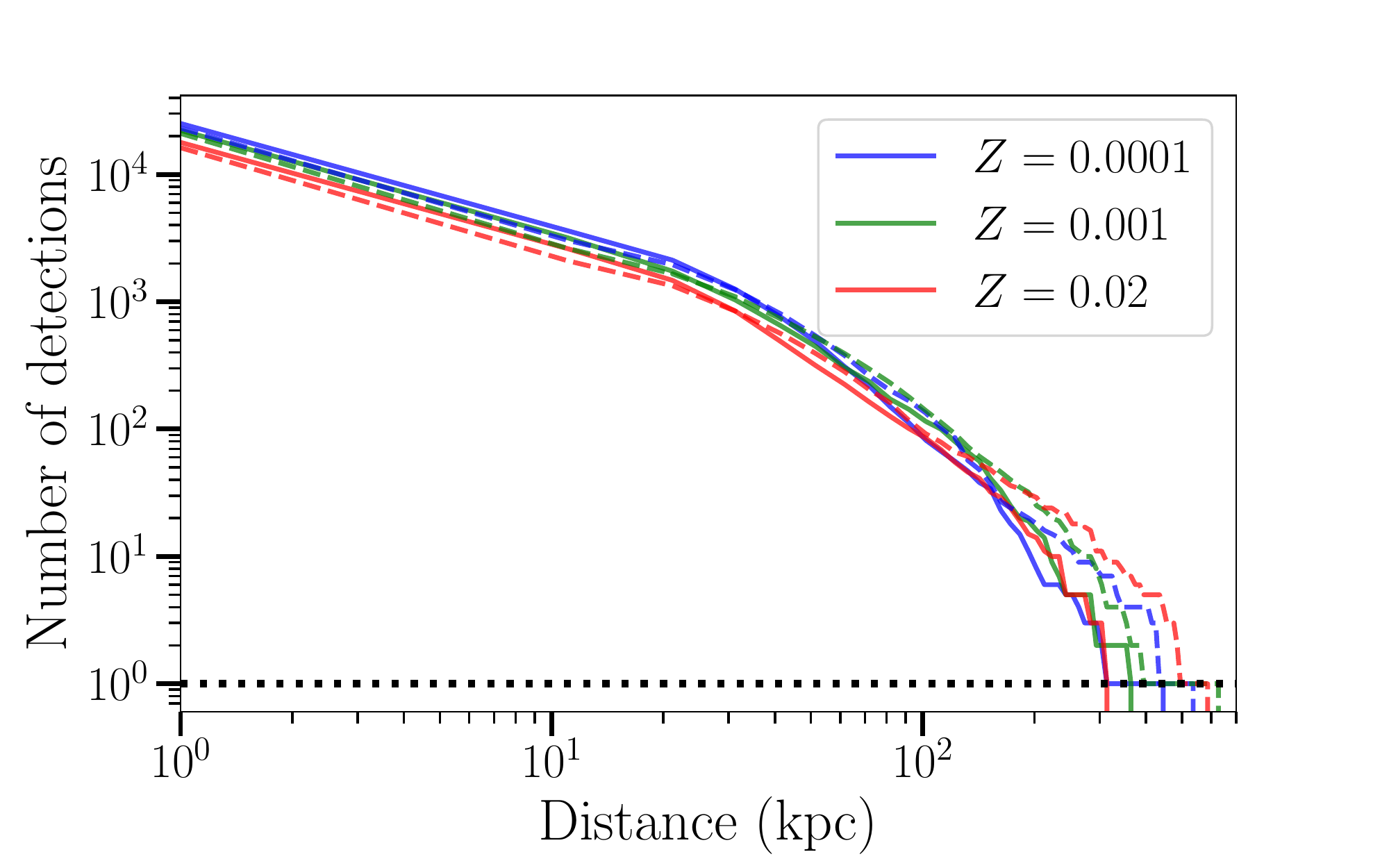}
	\caption{Number of detections as a function of distance for a satellite of $M_\star=10^{10}\,M_\odot$ with an exponentially decreasing SFH. Colours correspond to different values of metallicity: $Z=0.0001$ in blue, $Z=0.001$ in green and $Z=0.02$ in red.  Solid and dashed lines represent respectively $\gamma \alpha$- and $\alpha \alpha$-CE models. The dotted line shows the single-detection threshold.} 
	\label{fig:nr_def_zce}
\end{figure}

The number of detectable sources weakly increases with decreasing metallicity. 
For Solar metallicity the number of DWDs decreases by  $\sim$20\% compared to the default model  ($Z=0.001$) while at low metallicity ($Z=0.0001$) the number of DWDs increases by $\sim10\%$ out to distances of 50\,kpc (beyond this the number of sources is too low to establish any trend). 
Figure~\ref{fig:nr_def_zce} shows that the number of detections depends only moderately on metallicity, with an overall variation in the predicted rates of less than a factor of 1.5 (see also Table~\ref{tab:results}). This is in stark contrast with the case of binary black holes mergers observable by LIGO where the metallicity impacts the formation rate by 1 - 3 orders of magnitude  \citep[e.g.][]{Bel10, Gia18, Nei19}.
In general, metallicity alters the evolution of a star by changing its radius, core mass, and strength of the stellar winds. In case of the DWD evolution, metallicity mainly influences the minimum mass for an (isolated) star to reach the white dwarf stage in a Hubble time. This increases from $0.81\,M_\odot$ at $Z=0.0001$ to $0.82\,M_\odot$ at $Z=0.001$, and up to $0.98\,M_\odot$ at $Z=0.02$.

We are neglecting potential correlations between metallicity and primordial binary fraction. The advent of large and homogeneously selected samples indicates an anti-correlation for close ($\lesssim$10\,au) low-mass binaries \citep[][and references therein]{Bad18, Elb19, Moe19}. 
For instance, \cite{Bad18} find that the multiplicity fraction of metal-poor stars ($Z\lesssim0.005$) is enhanced by a factor 2 - 3 compared to metal-rich stars ($Z\gtrsim0.02$). Additionally, \citet{spe18} found that the binary fraction is not constant across the Milky Way's satellites.
In particular, Draco and Ursa Minor presents binary fractions of 50\% and 78\% respectively.
Enhancing the binary fraction from $50\%$ to  
70\% (90\%) in our simulations causes an increase of the number of detectable DWDs  by a factor of  $\sim$1.3 (1.6).

\begin{figure}
	\centering
	 \includegraphics[width=0.8\columnwidth]{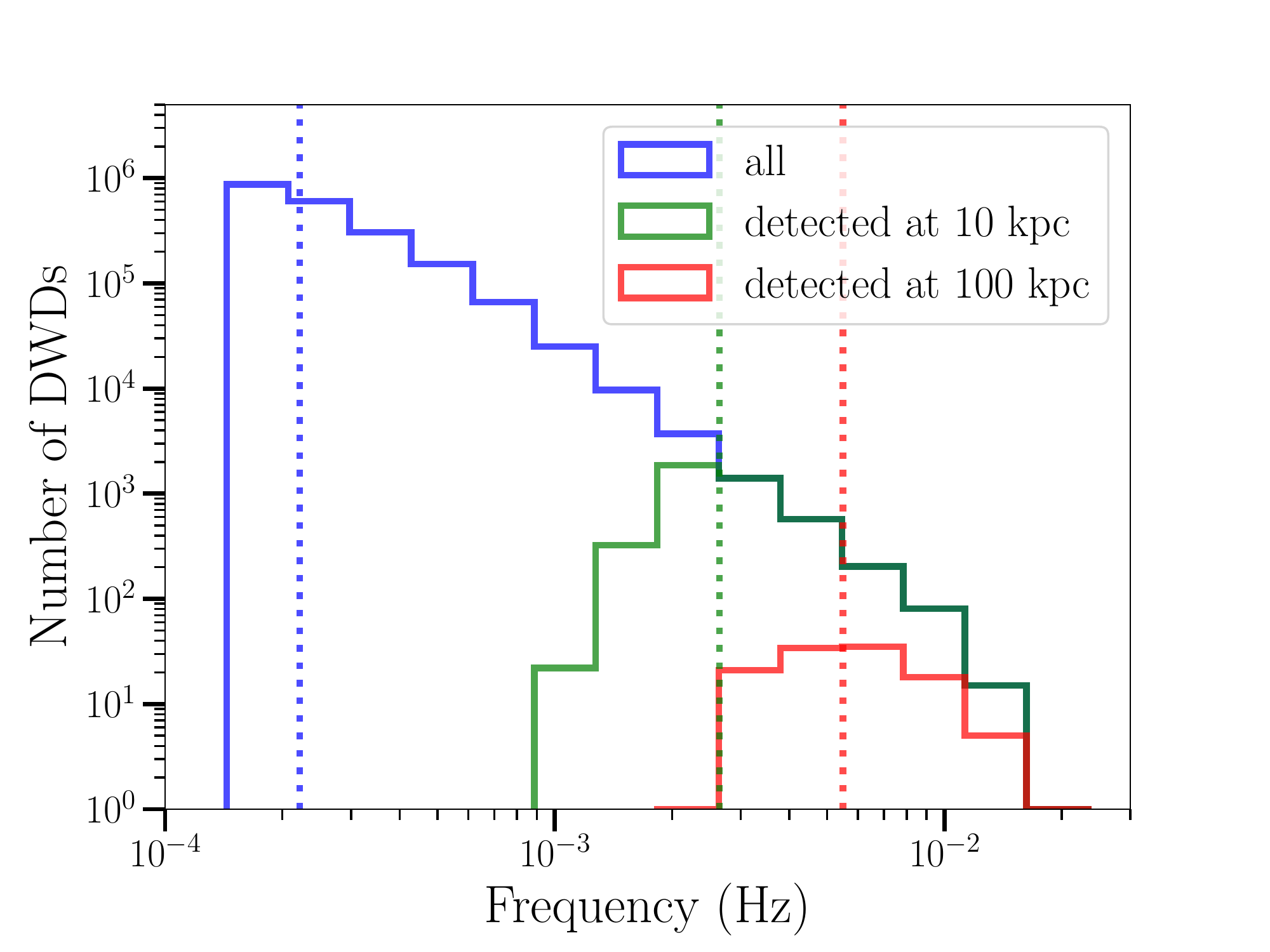} 
    \includegraphics[width=0.8\columnwidth]{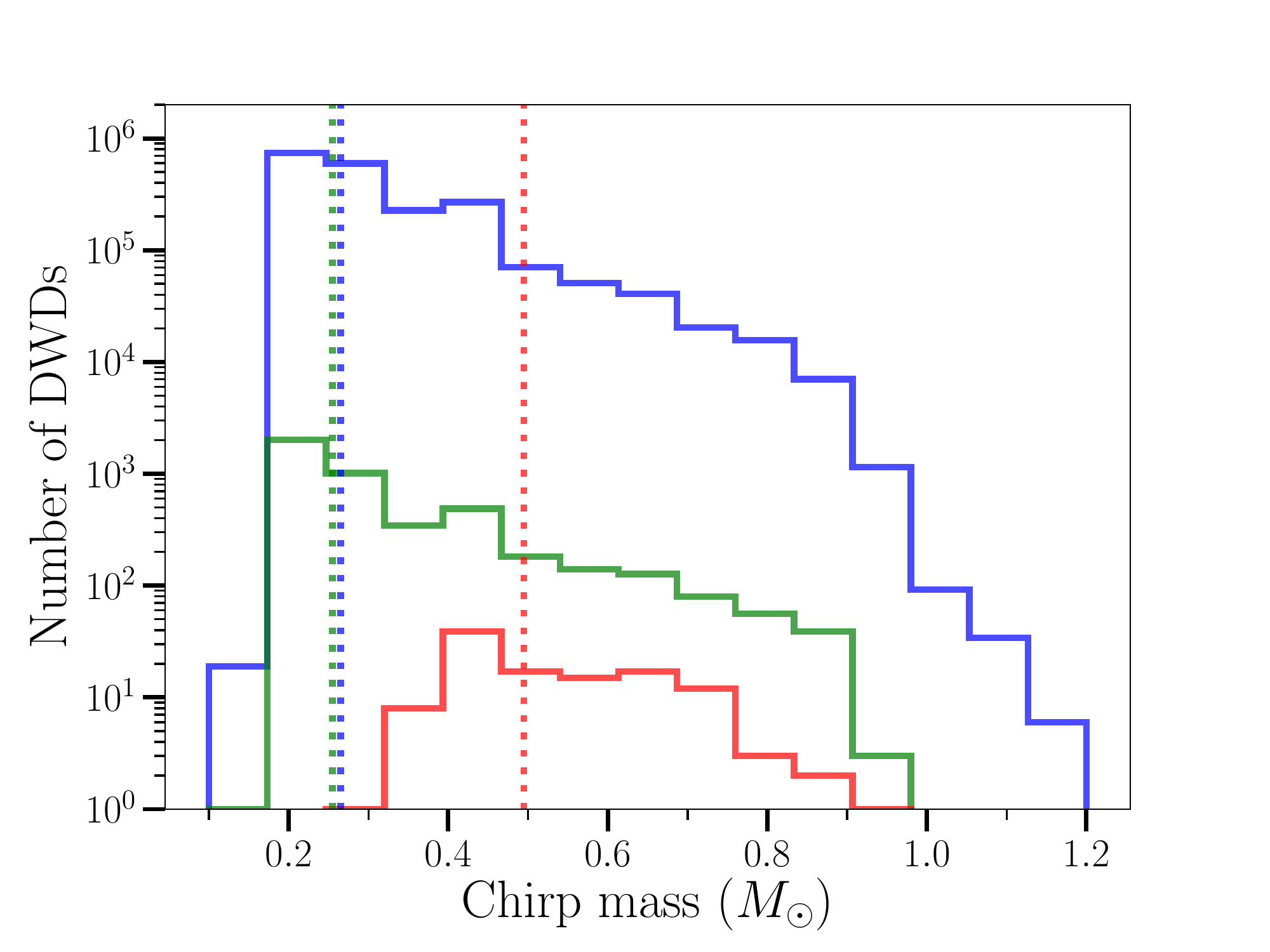}
    \includegraphics[width=0.8\columnwidth]{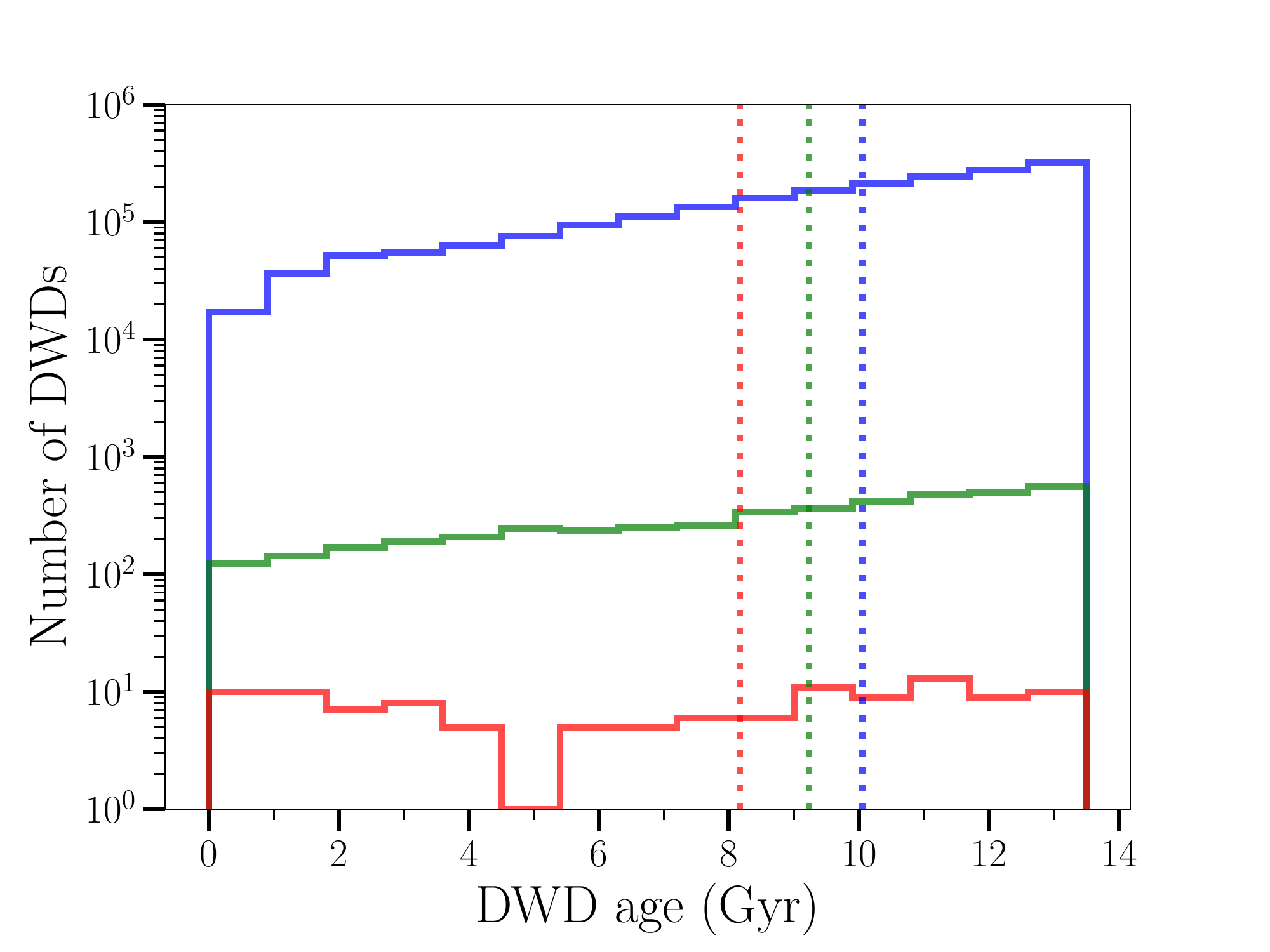}
	\caption{Distribution of frequencies, chirp masses and ages of DWDs in our fiducial satellite model (blue), of those detected by LISA at 10\,kpc (green) and 100\,kpc (red). Vertical coloured lines mark medians of the respective distributions.}
	\label{fig:properties}
\end{figure}

\begin{table}
\caption{Number of detectable DWDs hosted in  Milky Way satellites for nominal (4 yr) and extended (10 yr) mission duration derived using our fiducial exponentially declining SFH. Results using more realistic star formation models tuned for each satellite are reported in the text. Stellar masses and distances of satellites are adopted from \cite{mcc12}.}

\label{table:known_satellites}  
\begin{threeparttable}
\centering          
\begin{tabular}{l|cc|cc}
\toprule
\toprule
Name & $d\,$(kpc) & $M_\star (\times10^6\,$M$_{\odot}$)  & 4\,yr & 10\,yr \\
\midrule

LMC & 51 & 1500 & 70 & 150 \\
SMC & 64 & 460 & 15 & 30 \\
Sagittarius & 26 & 21 & 3 & 9 \\
Fornax & 147 & 20 & 0.1 & 0.3 \\
Sculptor & 86 & 2.3 & 0.04 & 0.1 \\
\bottomrule
\end{tabular}
\begin{tablenotes}
\small
\item {\bf Notes.} Results using more realistic star formation models tuned for each satellite are reported in the text. Stellar masses and distances of satellites are adopted from \cite{mcc12}.
\end{tablenotes}
\end{threeparttable}    
\end{table}

\subsection{Source properties}

Figure~\ref{fig:properties} illustrates the frequencies, chirp masses and ages (i.e. how long since the ZAMS) of detectable DWDs in our fiducial satellite of $10^{10}\,M_\odot$, 13.5\,Gyr with the exponential SFH for a LISA mission of 4\,yr.
The blue line represents the overall population, whereas the green and red lines represent the populations detected respectively at 10\,kpc (typical distances for stars in the Milky Way inner halo) and 100\,kpc (typical distances for the outer halo). 
Sources with higher frequencies and higher chirp mass remain detectable with increasing distance. In particular, the median value of the frequency distribution shifts from $\sim3$\,mHz at 10\,kpc to $\sim6$
\,mHz at 100\,kpc, while the median value of the chirp mass increases from $0.25\,M_\odot$ to $0.5\,M_\odot$ for the same values of $d$.

In our fiducial model, the number of DWDs increases with increasing age (blue solid line in the bottom panel of Fig.~\ref{fig:properties}) as a consequence of the adopted SFH and typical DWD formation timescales.  
The age distributions of binaries detected at 10\,kpc (in green) does not show a strong selection effect compared to the overall population, although the median is shifted towards smaller ages. 
The selection effect is much stronger for the DWDs at 100\,kpc  (in red),  shifting the median age to 8\,Gyr. These are CO+CO and CO+He DWDs with higher chirp masses (see Fig.~\ref{fig:pop0}).

\subsection{Known satellites} \label{sec:known_sats}

Ultra-faint dwarfs are unlikely to host detectable DWDs, therefore in this section we consider only classical and bright dwarfs with $M_\star \ge 10^6\,M_\odot$. We list stellar masses and distances of  known satellites from \citet{mcc12} and report the number of LISA detections for our fiducial model in Table~\ref{table:known_satellites}. Of all known satellites, only Sagittarius, Fornax, Sculptor, and the Magellanic Clouds host detectable LISA sources. 
For all the other satellites the probability of hosting LISA detections is $\lesssim$ 1\%.

Dwarf galaxies show a large variation of SFHs even within the same class and morphological type (cf. Sec.~\ref{sec:sfh}).
The SFH of the Sagittarius dwarf galaxy could be modelled by two events: a first one forming 50\% of the total stellar mass >12\,Gyr ago, and the other forming the remaining 50\% < 4\,Gyr ago \citep{wei14}. 
When  adopting this more appropriate star formation scenario, the number of detections in Sagittarius increases to 10 DWDs for a 4\,yr  mission lifetime compared to 3 reported in Table~\ref{table:known_satellites}. 
For the Fornax galaxy \citet{wei14} reports a constant star formation history, that based on our binary population modelling predicts 0.2 detectable DWDs (cf. Fig.~\ref{fig:detections}).
On the other hand, Sculptor is perhaps better described by a single star formation event that occurred $\sim12\,$Gyr ago, and thus our estimates are likely to be too optimistic for its case.
Both Magellanic Clouds have recently ($< 1\,$Gyr ago) experienced relevant star formation events \citep[e.g. ][]{str19,har09}.
Our declining star formation history models are likely to underestimate the number of detectable sources. 
For example, assuming a more optimistic constant SFH yields 100 (260) and 25 (55) detections respectively for the LMC and SMC assuming 4 (10)\,yr of LISA data. These  brightest satellites will be targeted separately in a forthcoming paper by \citet{kei20}.

In the companion paper by \citet[][figure 1]{roe20} we show the distribution of the sky distribution of known satellites from \citet{sim19,mcc12} highlighting those with detectable DWDs.

\section{Discussion} \label{sec:discussion}
Dwarf satellites with masses $10^6 - 10^{9}\,M_\odot$ host DWDs radiating GWs at frequencies $\gtrsim 3\,$mHz that are detectable by LISA out to distances of $10 - 300$\,kpc.
Specifically, we find that ultra-faint dwarfs will not host DWDs because (i) their total stellar mass is too low and (ii) they form stars only at early times.  
Classical dwarfs can be detected at distances from a few to several tens of kiloparsecs only if they experienced a significant star formation at recent times. 
However, bright dwarfs with $M_\star > 10^8\,M_\odot$ like the Magellanic Clouds can be detected at up to a few hundred kiloparsecs.
\citet{roe20} shows that these DWDs will not only be well localised on the sky ($< 10\, {\rm deg}^2$ for $\geq5$~mHz), but their distances will be measured  with precision of $10-40$\%. 
Shining bright in GWs, DWDs can be used as mass tracers at large galactocentric distances further enabling the characterisation of the Milky Way outer halo.

\subsection{Comparison with electromagnetic mass tracers}

Current EM mass tracers in the outer Milky Way typically represent collections of stars with a similar age, chemical composition and luminosity. Their 3D spatial distribution --sometimes also in combination with kinematics-- is required to map the baryonic matter distribution in our Galaxy and in the entire Local Group. 
The most common and abundant stellar  tracers present in optical surveys are main-sequence turn-off stars, blue horizontal branch stars, and M-giants stars \citep[for a review see][]{bel13}.
These stellar classes are numerous, luminous, and can be selected with low levels of contamination. 

Other commonly used mass tracers are variable stars such as RR Lyrae (RRL) and Cepheids.
Variables have a well-determined relation between period and absolute luminosity and thus serve as standard candles to measure distances \citep[e.g. ][]{cze18}.
Interestingly, almost all dwarfs in the Local Group that have been studied so far contain at least one RRL \citep{bak15}. 
This makes searches for stellar systems co-distant with RRLs a plausible mean to investigate  sub-structures with low surface brightness \citep[e.g.][]{ses14,bak15,tor19}.

In comparison, GW emitters  such as DWDs do not present contaminants. 
Their GW signals can potentially be used to trace dwarf satellites all the way out to the virial radius of the Galaxy. 
Thus, DWDs are competitive with RRL stars that can be found up to $\sim100\,$kpc in the {\it Gaia} data \citep{holl18}.

\cite{new18} recently assessed the current   observational limitations on the number of Milky Way dwarf satellites, and presented predictions for future optical surveys. Based on a number of cosmological simulations, they estimated that $124^{+40}_{-27}$ dwarfs galaxies with $M_V \sim 0$ should be present within galactocentric distances of $\lesssim300$\,kpc. Authors find that only half of predicted systems can be detected using the Rubin Observatory because of dust extinction and sky coverage limitations.

Although DWD observations do not suffer from any of these issues, they are much rarer compared to any of the aforementioned stellar mass tracers. The number of DWDs with $f\lesssim 3\,$mHz in a satellite is directly proportional to its total stellar mass, and quickly drops to zero for satellites with $M_\star < 10^6\,M_\odot$ and no recent star formation. 
On the other hand, \citet{roe20} show that DWDs with $f\gtrsim 3\,$mHz  will also have measurable frequency evolution, allowing distances to satellites to be measured.

\subsection{Weighing the satellites with GWs}

The number of LISA detections per satellite strongly depends on its total stellar mass (cf. Fig.~\ref{fig:detections}). 
If a group of co-distant GW sources can be identified in the LISA data, one can then reverse-engineer our modelling process to get the mass of the (known or unknown) satellite. 
We stress that GW detections yield the original stellar mass of a satellite including the contribution of evolved stars that are no longer visible through light. This is in contrast to  stellar masses derived from satellites' EM brightness, sensitive to the mass enclosed in bright stars.
Those kind of estimates are typically made by modelling the brightness of a satellite and applying age-dependent $L/M$ ratios from stellar calculations which must  necessarily adopt an IMF. This method is only sensitive to significant derivations from the nominal Kroupa-type IMF.

As an alternative, we use the IMF derived by~\cite{1979ApJS...41..513M} which is flat below $1\,M_\odot$ and is characterised by a higher mean mass of $0.65\,M_\odot$ compared to $0.49\,M_\odot$ for the Kroupa IMF. 
This IMF favours typical DWD progenitors. 
In comparison, the fiducial Kroupa IMF is more bottom-heavy and generates a higher number of low-mass stars that will need more than a Hubble time to turn into white dwarfs. 
We find that the Miller-Scalo IMF generates $\sim$10\%  more DWDs  per $3 \times 10^7$\,$M_\odot$, and, in particular, produces $\sim$30\% more DWDs in the LISA band. This strong increase is largely due to the presence of more massive CO+CO DWDs.
However when evolving the population to the age of the satellite  (Section~\ref{sec:sfh}), these more massive short-lived CO+CO DWDs merge within a few Gyr, such that predominantly the low-frequency binaries ($<1\,$mHz) remain. As already discussed, these binaries are hardly detectable beyond $\sim10$\,kpc. 
On the contrary, the Kroupa IMF generates more low-mass progenitors which take longer to turn into DWDs and evolve to mHz frequencies.
Consequently, assuming a bottom-heavy IMF such as the Kroupa-IMF  our simulations predicts more detections in satellites of intermediate and old age which are detectable at larger galactocentric distances. 

As a quantitative example, we model the Sagittarius dwarf galaxy adopting a double-burst SFH and the age of 13.5\,Gyr as described in Section~\ref{sec:known_sats}. We find that the Miller-Scalo IMF predicts a similar number of LISA detections as the fiducial Kroupa IMF. 
However, DWDs in the former case with lower GW frequencies are difficult to identify as extra-galactic against the Galactic confusion foreground, due to large errors on the distance at low frequencies \citep{roe20}.
From this example we conclude that we can identify a relatively low-mass satellite in the Milky Way halo, provided its population originated from a more bottom-heavy  IMF.

\section{Conclusions} \label{sec:conclusions}

LISA can detect short-period DWDs beyond our galaxy, potentially reaching the outskirts of the Local Group. In this paper we assess what properties qualify a dwarf satellite galaxy as a host for LISA sources. We use a  treatment tailored on individual Milky Way satellites. Complementary predictions where presented by \citet{Lamberts19} using cosmological simulations. 
We use binary population synthesis to produce samples of DWDs for  a putative galaxy with fixed binary fraction, metallicity, IMF, age and an analytic star formation model. Simulation results are then re-scaled to the mass of known Milky Way satellites.
Our simple approach allows us to vary this set of assumptions and  determine which region of the parameter space is optimal for hosting LISA sources. Specifically, our finding articulates as follows. 
\begin{itemize}
    \item The number of DWDs strongly depends on the satellite's total stellar mass. This limits LISA detections to satellites with $M_\star >10^6\,M_\odot$.
    
    \item Both age and SFH influence the DWD orbital period distribution, and consequently  number of binaries with $f \gtrsim 3\,$mHz that are detectable at large distances. Because in young and star-forming satellites the birthrate of DWDs peaks at early times ($\sim 1$\,Gyr), these systems are expected to host more DWDs emitting GWs in the LISA band.
        
    \item Metallicity has a limited effect on DWD populations, with the total number of binaries increasing for decreasing metallicities. 
    
    \item Only DWDs with frequencies $\gtrsim 3\,$mHz can be detected beyond the Milky Way stellar disc. This implies that all most of extra-galactic LISA detections will have measurable frequency derivative allowing to pin down the distance to the source with a relative error of $\sim 30\%$ \citep{roe20}. These DWDs can be used as mass tracers in the outer Milky Way much like Cepheids and RRL stars.
    
    \item Of the known satellites only  Sagittarius,  Fornax,  Sculptor  and  the  Magellanic  Clouds host  detectable population of LISA  sources.
    
    \item  If a suitable number of detections are identified within a satellite, inference on the satellite's distance one can be used to  estimate its total stellar mass.

\end{itemize}

As an all-sky survey that does not suffer from contamination and dust extinction, LISA can detect known Milky Way dwarf satellites and potentially discover new ones through populations invisible to EM instruments. 
Properties of these populations will inform us on the formation history of the Milky Way and its environs and provide unique contribution to Galactic archaeology and near-field cosmology.

\begin{acknowledgements}
VK thanks J. Dalcantan for inspiring this study and K. Duncan for useful discussions on the observed electromagnetic properties of stellar populations.
VK and ST acknowledge support from the Netherlands Research Council NWO (respectively Rubicon 019.183EN.015 and VENI 639.041.645 grants). 
FV acknowledges support from the European Research Council Consolidator Grant
funding scheme (project ASTEROCHRONOMETRY, G.A. No.~772293) and the support of a Fellowship from the Center for Cosmology and AstroParticle Physics at The Ohio State University.
DG is supported by Leverhulme Trust Grant No. RPG-2019-350. AV acknowledges support from the Royal Society and the Wolfson Foundation. Computational work was performed on the University of Birmingham BlueBEAR cluster, the Athena cluster at HPC Midlands+ funded by EPSRC Grant No. EP/P020232/1, and the Maryland Advanced Research Computing Center (MARCC).
\\
This research made use of open source \texttt{python} package \texttt{smpy}, developed by K. Duncan and available at \url{https://github.com/dunkenj/smpy}, \texttt{NumPy, SciPy, PyGaia python} packages and matplotlib \texttt{python} library. 
\end{acknowledgements}

\bibliographystyle{aa} 
\bibliography{bibtex_wd}

\end{document}